\documentclass[twocolumn,showpacs,preprintnumbers,amsmath,amssymb]{revtex4}

\usepackage{graphicx}
\usepackage{dcolumn}
\usepackage{bm}
\usepackage{amssymb}

\begin{document}

\preprint{Aste-Valbusa, Ripples. 6.1, \today}

\title{ Ripples and Ripples: from Sandy Deserts to Ion-Sputtered Surfaces}

\author{ T. Aste$^{1,2}$}
\email{tomaso.aste@anu.edu.au }
\author{U. Valbusa$^2$ }
\affiliation{$^1$ Applied Mathematics, RSPHYSSE, ANU, 0200 Canberra ACT, Australia }
\affiliation{$^2$ INFM-Dipartimento di Fisica, Universit\`a di Genova, via Dodecaneso 33, 16146 Genova, Italy}


%

%
\begin{abstract}
We study the morphological evolution of surfaces during ion sputtering and we compare their dynamical roughening with aeolian ripple formation in sandy deserts. 
We show that the two phenomena can be described within the same theoretical framework. 
This approach explains the different dynamical behaviors experimentally observed in metals or in semiconductors and amorphous systems. 
In the case of ion erosion, we find exponential growth at constant wavelength up to a critical roughness $W_c$. 
Whereas, in metals, by introducing the contribution of the Erlich-Schwoebel barrier, we find a transition from an exponential growth to a power law evolution.
\end{abstract}

\maketitle

\section{Introduction}

When an ion hits a surface liberates locally a large amount of energy that melts a region of the solid immediately below.
For geometrical reasons, the sputtering effect depends on the surface-curvature: the energy concentrates on regions of positive curvature and this favorites the excavation of valleys and the growth of hills.
On the other hand, thermal diffusion and surface tension tend to smoothen the irregularities by flattening the surface.
It has been observed that under the combined action of these mechanisms the surface tends to create spontaneously ripples \cite{Ekl91,Krim93,Chra94,Valbusa02,Buat00,Rusp98,Rusp98a,Rusp97}. 
In nature, ripples are commonly observed in sandy-deserts as the result of dynamical instability of the sand surface under the action of a sufficiently strong wind \cite{Bagnold}.
In this case, the formation of ripples is commonly associated with the effect produced by some grains that are lifted from the sand-bed and accelerated by the wind.
These grains, when re-impact with the bed, splash up a number of other grains. Most of these grains return to the bed leading to a local rearrangement, whereas some other are accelerated by the wind and impact again on the bed after a certain `saltation' length.
In the literature, many studies have been devoted to understanding the mechanism of ripple formation \cite{Andersen02,Andersen01,Kurtze00,Hoyle99,Prigozhin99,Werner93}.
In particular, an hydrodynamical model for aeolian ripple formation, based on a continuum dynamical description with two species of grains (immobile and rolling grains), was proposed with success by Bouchaud et al. \cite{Bouch95,Terz98,Val99,Csa00}. 
The main ingredient of such a model is a bilinear differential equation, for the population of the two species of grains, which shows the instability of a flat bed against ripple formation.

In this paper we show that the same reasoning which has been used to describe the sand ripples formation in deserts, applied to the studies of dynamical surface roughening, leads to an accurate description of the morphogenesis and evolution of ripples on crystal and amorphous surfaces during ion sputtering.
The present approach contains not only the Bradley-Harper approach \cite{Bra88,Par99} (based on a Kardar-Parisi-Zhang type equation \cite{KPZ}), but it is also able to describe some of the crucial experimental features recently observed in these systems \cite{Rusp98}.
In particular, by means of this approach we can explain the two distinct dynamical behaviors experimentally observed in amorphous/semiconductors systems and in metals \cite{Valbusa02}. 
In the first case (amorphous/semiconductors) we find that the ripples growth exponentially fast at constant wavelength $\hat \lambda$ up to a critical roughening $W_c$ at which the growing process interrupts.
On the other hand, in metals (when the Erlich-Schwoebel barrier is active), we find a transition between an initial exponential to a slower power-law growth of the roughness (the root mean square of the height profile).
In this regime the ripple-wavelength tends also to growth with time.

\begin{figure}
\hspace{+2cm}
\includegraphics[width=7cm]{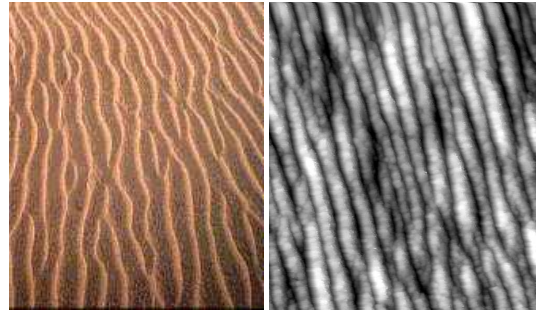}
\caption{\label{f.0} 
Ripples on sand (Gobi desert) and Ripples on surfaces (Ag under ion sputtering).}
\end{figure}

\section{Particle mobility and Ripple dynamics}

When the surface of a solid is taken under ion sputtering some atoms in the proximity of the surface receive energy from the sputtered ions and pass from a bounded - \emph{`immobile'} - solid state to a \emph{`mobile'} melted unbounded state.
The opposite mechanism is also allowed: some \emph{mobile} atoms can gain in energy by becoming \emph{immobile} and bounding in a given position in the solid.
A certain fraction of atoms might also be dispersed into the atmosphere. 
Let us call $h({\mathbf r },t)$ the height of surface profile made of immobile -bounded- atoms and call $R({\mathbf r },t)$ the height of mobile -melted- atoms.
In analogy with the theory developed to explain the dynamical evolutions of dunes in deserts \cite{Bouch95,Terz98,Val99,Csa00}, we describe the mechanisms of excavation, exchange between mobile and immobile atoms and surface displacement of mobile atoms in term of the following differential equation:
\begin{eqnarray} \label{E1}
\frac{\partial h}{\partial t} 
&=&  - \Gamma(R,h)_{ex} + \Gamma(R,h)_{ad} 
\nonumber \\
\frac{\partial R}{\partial t} 
&=& - {\mathbf \nabla} {\mathbf J}(R,h) + (1-\phi) \Gamma(R,h)_{ex} -\Gamma(R,h)_{ad} \;\;\;.
\end{eqnarray}
Where $\Gamma(R,h)_{ex}$ is the rate of atoms that are excavated under the action of the sputtering, and $(1-\phi)$ is the part of them that pass from immobile to 
mobile, whereas $\phi$ is the fraction that is dispersed into the atmosphere. 

Let us now write in details the various terms contained in Eq.\ref{E1}.

\subsection{Excavation}
The excavation effect must clearly depend on the number and velocity of the sputtered ions (i.e. its flux), but also the local shape and orientation of the surface might play an important role.
Indeed, the energy transmitted by the impacting ions concentrate more in regions of the surface with positive curvature. 
Moreover, part of the surface facing the flux are likely to experience a different erosion respect to others which are less exposed to the flux.
Crystalline orientation and anisotropies might be also taken into account.
We can write:
\begin{equation}\label{Gex}
\Gamma(R,h)_{ex} = 
\eta 
( 
1 
+ {\mathbf  a }    {\mathbf \nabla} h 
+ { b }    {\mathbf \nabla}^2 h  
) \;\;\; ;
\end{equation}
here $\eta$ is the sputtering flux, whereas ${\mathbf a}$ and ${ b }$ are respectively associated with the flux-direction-dependent and with the curvature-dependent sputtering erosion.

\subsection{Adsorption}
The rate of adsorption of mobile atoms into immobile solid positions must be dependent on the quantity of mobile atoms in a given spatial position.
Similarly to the excavation process, the adsorption is also dependent on the local curvature and orientation. 
We can write:
\begin{equation}\label{Gad}
\Gamma(R,h)_{ad} =  
R 
( 
\gamma 
+ {\mathbf  c }    {\mathbf \nabla} h 
+ { d }    {\mathbf \nabla}^2 h  
) \;\;\;,
\end{equation}
where the parameter $\gamma$  is the recombination rate and ${\mathbf c}$ and ${ d }$  are associated to the different probabilities of recombination in relation with the local orientation and shape of the surface. 

Note that Eqs. \ref{Gex}, \ref{Gad} contains the same terms as the ones proposed in the literature for the formation of aeolian dunes in the so-called hydrodynamical model \cite{Bouch95,Terz98,Val99,Csa00,Mak00}.
Indeed, in deserts, sand grains are lifted from the sand-bed and readsorbed into it with a probability which is dependent on the local shape and orientation of the dunes.
Eqs. \ref{Gex}, \ref{Gad} represent the simplest analytical expressions which formally take into account these shape and orientation dependences.
In the search for simple explanations, such equations are therefore rather universal.

\subsection{Mobility}
Mobile atoms will move on the surface, and the quantity ${\mathbf J}(R,h) $ in Eq.\ref{E1} is the `current' of these atoms. 
In surface growth, there are two main mechanisms that are commonly indicated as the responsible for the surface mobility of atoms \cite{Bar95}.
The first is a current, driven by the variations of the local chemical potential, which tends to smoothen the surface asperity moving atoms from hills to valleys. The second is the current induced by the Erlich-Schwoebel barrier which -on the contrary- moves atoms uphill.
In addiction to these main mechanisms we might also have to take into account a drift velocity and a random thermal diffusion, obtaining:
\begin{equation} \label{J1}
{\mathbf J}(R,h)  = 
{\mathcal K} R {\mathbf \nabla}({\mathbf \nabla}^2 h)
+ s R  \frac{ {\mathbf \nabla} h }{1+ ( \alpha_d {\mathbf \nabla} h)^2}
+
{\mathbf v } R 
- {\mathcal D } {\mathbf \nabla } R 
  \;\;\;.
\end{equation}
In this equation, the first term describes a deterministic diffusion driven by the variations of the chemical potential which depends on the local shape of the surface; the second term is associated with the uphill current due to the Erlich-Schwoebel barrier and $\alpha_d$ is a constant associated with the characteristic length.
The quantity ${\mathbf v }$ is a drift velocity of the mobile atoms on the surface, whereas ${\mathcal D }$ is the dispersion constant associated with the random thermal motion.

Note that Eq.\ref{J1} is substantially different from the one proposed in the literature to describe ripples in granular media \cite{Bouch95,Terz98,Val99,Csa92,Csa93}. 
Here the current is supposed to be dependent on the local shape and orientation of the surface (the $ h({\mathbf r },t)$ profile). 
The equations describing sand deserts can be retrieved from Eq.\ref{J1} by imposing ${\mathcal K}=0$ and $s=0$, but -on the contrary- in surface growth these two parameters are the leading terms of the equation and play the role of control parameters in the dynamics of ripple formation.
Nonetheless, these terms describe a rather simple dependence of the dynamic of particles on a surface on the geometrical shape of the surface itself.
Again, in our seek for universality, we expect that similar terms can be profitably introduced in the context of aeolian sand ripples in order to describe specific phenomena (associated, for instance, with packing properties \cite{Aste00} or granular flow \cite{ Komatsu01}) which relate the current of grains with the dune-shapes.

It should be noted that the factors ${\mathbf a}$, ${\mathbf c}$ and ${\mathbf v}$ in Eqs.\ref{Gex},\ref{Gad} and \ref{J1} are \emph{vectors} (i.e. they have -in general- different components in the two horizontal directions).
Indeed, crystal surfaces are in general anisotropic and therefore one must take into account the dependence of the parameters on the relative orientation of the crystal-surface and the sputtering direction.

\begin{figure}
\hspace{+2cm}
\includegraphics[width=8.5cm]{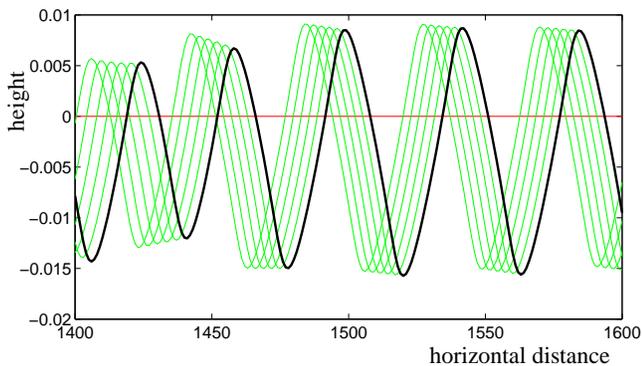}
\caption{\label{f.1} 
Numerical solutions of Eq.\ref{E1} at various times indicate that under the action of ion sputtering the surface develops an instability which leads to the formation of ripples with a well defined characteristic wavelength.
In the figure the black-tick line is the final surface-profile, whereas the tinnier-gray lines are some profiles at previous times.}
\end{figure}

\section{Dispersion Relation}

A trivial solution of Eq.\ref{E1} can be written for a completely flat surface: $h({\mathbf r },t)=h_0(t)$ and $R({\mathbf r },t)=R_0$.
In this case, we obtain $R_0 = (1-\phi) \eta/ \gamma$ and $h_0(t)=-\phi \eta t + const.$.
This describes a surface that rests flat and it is eroded with a speed equal to $\phi \eta$.
But this behavior is only hypothetical since -in general- the dynamics of the surface-profile presents instabilities against spontaneous roughening and therefore its evolution is more complex.
For instance, a numerical solution of Eq.\ref{E1}, is shown in Fig.\ref{f.1} (for the 1-dimensional case).
We observe that, in a certain range of the parameters, the surface is unstable and periodic ripples are formed spontaneously.

\subsection{Stability analysis}

In order to infer indications about the amplification or the smoothing of small perturbations and to deduce an analytical expression for the ripples wave-length  at their beginning, we performe a stability analysis on Eq.\ref{E1}.
For this purpose we assume that the surface-profile is made by the combination of a flat term plus a rough part: 
\begin{eqnarray}\label{R1h1}
R({\mathbf r },t) &=& R_0 + R_1({\mathbf r },t) \nonumber \\
h(x,t) &=& h_0(t) + h_1({\mathbf r },t) \;\;\;,
\end{eqnarray}
with $ R_1({\mathbf r },t) = \hat R_1 \exp(i \omega t + i {\mathbf k } {\mathbf r })$ and $ h_1({\mathbf r },t) = \hat h_1 \exp(i \omega t + i {\mathbf k } {\mathbf r })$. 
We substitute these quantities into Eq.\ref{E1} and linearize the equation by taking only the first order in $R_1$ and $h_1$.
A Fourier analysis (see Appendix \ref{A1}) shows that such a linearized equation admits solutions when the frequencies $\omega$ and the wave vectors ${\mathbf k}$ satisfy:
\begin{eqnarray} \label{Deter}
& &\left[ 
i \omega + 
\gamma 
+ i {\mathbf k } {\mathbf v } 
+  k^2 {\mathcal D } 
\right] \cdot
\nonumber \\
& &\left\{
i \omega 
+  i {\mathbf k } \left[ {\mathbf v }_1 - (1-\phi) {\mathbf v }_2 \right]
- k^2 \left[ {\mathcal D }_1 -(1-\phi) {\mathcal D }_2 \right]
\right\}-
\nonumber \\
& &
\gamma
(1-\phi)
\!
\left[
\!
i {\mathbf k } ({\mathbf v }_1 
\! - \!{\mathbf v }_2) 
\! - \! k^2  \left({\mathcal D }_1 
\! - \! {\mathcal D }_2 - s_1 \right)
\! - \! k^4 {\mathcal K }_1
\right] = 0
\;  ;
\nonumber \\
\end{eqnarray}
where, to simplify the equations, we have introduced the following notation: 
\begin{center}
\begin{tabular}{lll}
${\mathbf v}_1 = \eta {\mathbf a}$ 
\hspace{1.cm} &  ${\mathbf v}_2 = \eta {\mathbf c}/{\gamma}$ \\
${\mathcal D }_1 = \eta b$
& ${\mathcal D }_2 = \eta d/{\gamma}$ \\
$s_1 = \eta s/{\gamma}$
& ${\mathcal K }_1 = \eta {\mathcal K }/\gamma$ 
\end{tabular}
\end{center}
Equation \ref{Deter} establishes a \emph{dispersion relation} $\omega({\mathbf k})$ that is a complex function with two branches corresponding to the solutions of the quadratic Equation \ref{Deter}.

\section{Surface Instabilities}

\begin{figure}
\hspace{+2cm}
\includegraphics[width=8.5cm]{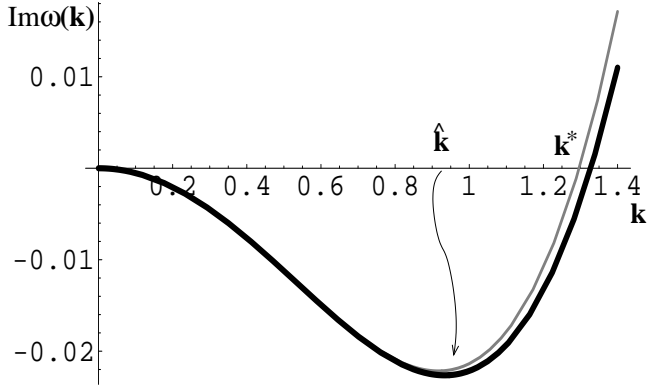}
\caption{\label{f.2} 
The imaginary part of the dispersion relation $Im(\omega)$ can assume negative values which are associated with the surface instability (arbitrary units) \cite{typical}.
The amplitude of modes with wavelengths $\lambda > 2\pi/k^*$ will grow exponentially fast.
The tick line is the imaginary part of the analytical solution of Eq.\ref{Deter}, whereas the tinny-gray line is the approximated expression (at the fourth order in $k$) obtained for small ion flux ($\eta$ small).  
}
\end{figure}

The kinetic growth of the surface instability is related to the immaginary part of $\omega({\mathbf k})$.
Indeed, $Im(\omega({\mathbf k}))$ corresponds to modes with amplitudes that change exponentially fast in time, and negative values correspond to unstable modes that increase with the time. 
We can therefore study $Im(\omega)$ from the solution of Eq.\ref{Deter} and search for the range of $k$ in which $Im(\omega)$ is negative.
The most unstable mode is the one that grows faster and it corresponds to the value of ${\mathbf k}$ at which $Im(\omega)$ reaches its most negative value (see Fig.\ref{f.2}).

The solution of Eq.\ref{Deter} for $Im(\omega)$, is 
\begin{eqnarray}\label{solut}
2 Im(\omega)_\pm 
& = &
\gamma +
 \Big[{\mathcal D } - {\mathcal D }_1 +(1-\phi) {\mathcal D }_2 \Big]k^2
\nonumber \\
& \pm &
\sqrt{\frac{\Delta_1 + (\Delta_1^2 + 4 \Delta_2^2)^{1/2}}{2}}
\end{eqnarray}
where we have
\begin{eqnarray}\label{delta1}
\Delta_1 &=& 
\gamma^2 - \Big\{ \Big[ {\mathbf v} - {\mathbf v}_1 +(1-\phi) {\mathbf v}_2\Big] {\mathbf k} \Big\}^2
\nonumber \\
&+&
2 \gamma \Big[ {\mathcal D } - (1-2\phi){\mathcal D }_1 + (1-\phi) {\mathcal D }_2 + 2 (1-\phi)s_1\Big] k^2
\nonumber \\
&+& 
\Big\{
	\Big[({\mathcal D } + {\mathcal D }_1 - (1-\phi){\mathcal D }_2\Big]^2 
	- 4 \gamma (1-\phi) {\mathcal K }_1
\Big\} k^4
\nonumber \\
\end{eqnarray}
and 
\begin{eqnarray}\label{delta12}
\Delta_2 &=& 
\gamma \Big[{\mathbf v}+ (1-2\phi){\mathbf v}_1-(1-\phi){\mathbf v}_2\Big] {\mathbf k }
\nonumber \\
&+&
\Big[{\mathcal D } + {\mathcal D }_1- (1-\phi){\mathcal D }_2\Big] 
\Big[{\mathbf v}-{\mathbf v}_1+(1-\phi){\mathbf v}_2 \Big]{\mathbf k }^3 \;\;.
\nonumber \\
\end{eqnarray}

Let us first observe that in absence of sputtering (i.e. when $\eta = 0$ and therefore, ${\mathbf v }_1=0$, ${\mathbf v }_2=0$, 
${\mathcal D }_1=0$, ${\mathcal D }_2=0$,
$s_1=0$, ${\mathcal K }_1=0$) the solutions of Eq.\ref{Deter} are $\omega({\mathbf k}) =0$ and $\omega({\mathbf k}) = - {\mathbf k } {\mathbf v } + i (\gamma + k^2 {\mathcal D })$.
In this case, the imaginary part of $\omega({\mathbf k})$ is non-negative, therefore we -correctly- expect no spontaneous corrugation of the surface.
On the contrary, when the sputtering is active ($\eta \not= 0$), the immaginary part of $\omega({\mathbf k})$ can assume negative values as shown in Fig.\ref{f.2} where a plot of $Im(\omega)_-$ is reported (along one direction of the vector ${\mathbf k}$).
As one can see, typically the branch $Im(\omega)_-$ takes negative values for $k$ between 0 and a critical value $k^*$ at which it passes the zero \cite{typical}.
The critical point $k^*$, fixes the minimal unstable wavelength.
We therefore expect to find unstable solutions associated with the formation and evolution of ripples with wavelengths $\lambda \ge \lambda^* = 2\pi/k^*$.

\section{Ripple wavelength}

Several analytical solutions of Eq.\ref{Deter} can be found in special cases which are discussed in Appendix \ref{A2}.
But the study of the surface instabilities can be highly simplified if we consider the first order effects when the sputtering flux $\eta$ is small.

\subsection{Approximate equation}

In the case of small spattering fluxes, the branch of $Im(\omega({\mathbf k}))$, with negative values can be approximated to:
\begin{equation}\label{dev}
Im(\omega)_-   
\simeq  
\frac{
P_1 k^6 + P_2 k^4 + P_3({\mathbf k})k^2 + P_4 k^2 + P_5({\mathbf k}) }{
{\mathcal D }^2 k^4 
+ 2 \gamma {\mathcal D } k^2 
+ ({\mathbf v  \mathbf k})^2
+\gamma^2  
}
\end{equation}
with
\begin{eqnarray}
&&P_1 = {\mathcal D } [ (1-\phi)\gamma {\mathcal K }_1 - {\mathcal D }({\mathcal D }_1 -(1-\phi) {\mathcal D }_2 )]
\nonumber \\
&&P_2 = (1-\phi) \gamma  [{\mathcal D }({\mathcal D }_2-s_1) + \gamma {\mathcal K }_1] - (1+\phi)\gamma {\mathcal D }{\mathcal D }_1  \nonumber \\
&&P_3({\mathbf k}) = - [{\mathcal D }_1 + (1-\phi) {\mathcal D}_2 ] ({\mathbf v}{\mathbf k})^2
\nonumber \\
&& P_4 = \gamma^2 [ (1-\phi) s_1 - \phi {\mathcal D }_1 ] 
\nonumber \\
&& P_5({\mathbf k}) = (1-\phi) \gamma ({\mathbf v}{\mathbf k}) [({\mathbf v}_1-{\mathbf v}_2){\mathbf k}] 
\end{eqnarray}

When $k$ is sufficiently small ( $k \ll \gamma/\eta$ ), we can develop  Eq.\ref{dev} at the 4$^{th}$ order obtaining:
\begin{equation}\label{4ord}
Im(\omega)_-  \simeq  A k^4 - B k^2 \;\;,
\end{equation} 
with
\begin{eqnarray}\label{AB}
A &=& (1-\phi) \Big[
{\mathcal K }_1 
+ (s_1 + {\mathcal D }_2 -{\mathcal D }_1) \frac{ \gamma {\mathcal D } + v^2}{\gamma^2}+
\nonumber \\
& & v (v_1-v_2) \frac{ 2 \gamma {\mathcal D } + v^2}{\gamma^3}
\Big]
\nonumber \\
B &=& \phi {\mathcal D }_1 + (1-\phi) \Big[ s_1  + \frac{ v (v_1-v_2)}{\gamma} \Big]\;\;.
\end{eqnarray}
Here $v$, $v_1$  and $v_2$ are respectivelly the components of ${\mathbf v}$, ${\mathbf v}_1$  and ${\mathbf v}_2$ in the direction parallel to ${\mathbf k}$).
(In Fig.\ref{f.2} a comparison between this approximate solution and the exact one is given.)

\subsection{Solutions}

The expected wavelength of the ripples is associated with the fastest growing mode, which corresponds to the value of ${\mathbf k}$ at which $Im(\omega)_-$ reaches its most negative point.
Here the minimum of $Im(\omega)$ is at 
\begin{equation}\label{khat}
\hat k = \sqrt{\frac{B}{2A}}\;\;\;.
\end{equation}
Therefore, at the beginning, the roughness will grow exponentially fast as $W \sim \exp(B^2 t/(4A))$ with associated ripple-wavelength at: \begin{equation}\label{wave}
\hat \lambda \sim 2 \pi \sqrt{\frac{2 A }{B }} \;\;\;\;.
\end{equation}

\begin{figure}
\hspace{+2cm}
\includegraphics[width=8.5cm]{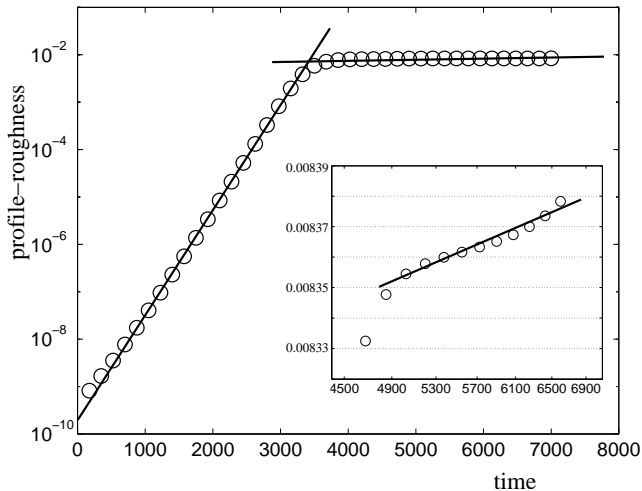}
\caption{\label{f.3} 
Evolution of the surface roughness (log-scale) v.s. time (linear scale) from numerical simulations (see Appendix \ref{A.sim}).
(Arbitrary units).
The insertion is the log-log plot of the last part of the evolution.
When the Erlich-Schwoebel barrier is active the dynamical evolution of ripples is characterized by an exponentially-fast growth at the beginning and then a `saturation' to a power-law growth (linear trend in log-log scale).
}
\end{figure}

\subsection{Some special cases}

Let us first observe that, when ${\mathcal K }_1 $, $s_1$ and $\phi = 0$, the ripple wavelength, given by Eq.\ref{wave}, coincides with the one found for sand dunes in deserts (see for instance \cite{Val99}). 
In our notation the `reptation length' is $l_0 = v/\gamma$, the `cut-off length' is $l_c = ({\mathcal D }_2-{\mathcal D }_1)/v$, whereas $v_1-v_2$ is the collective drift velocity of the dunes.
The approximations usually applied in this context \cite{Terz98,Val99}, imply: $l_c \gg \sqrt{{\mathcal D }/\gamma}$, and $\gamma l_c \gg v_1-v_2$.
Giving, from Eq.\ref{wave}
\begin{equation}
\hat \lambda \sim 2 \pi \sqrt{ \frac{2 v l_0 l_c }{v_1-v_2} } \;\;\;\;.
\end{equation}

Let us now consider the dynamical evolution of a surface under ion sputtering and in particular the case when the effect of the Erlich-Schwoebel barrier is not present (as for semiconductors and glasses). 
In this case, $s=0, s_1=0$ and we also expect that the drift velocity $v$ and the dispersion constant ${\mathcal D }$ are equal to zero or infinitesimally small.
Indeed, here the current of mobile atoms on the surface is mainly induced by the differences in the chemical potential. 
Under these assumptions, from Eq.\ref{wave}, the wavelength of the most unstable ripple is:
\begin{equation}\label{lk}
\hat \lambda \sim 2 \pi \sqrt{ \frac{2 {\mathcal K }}{\nu}} \;\;\;\;.
\end{equation}
where we called $\nu = \gamma b \phi/(1-\phi) $, a quantity which plays the role of an effective surface tension.
Note that Equation \ref{lk} is the same result as from the Bradley and Harper theory \cite{Bra88,Par99,Bar95,Cuer95,Gill01}.

When the Erlich-Schwoebel barriers are active ($s,s_1 \not= 0$), effects can be observed on the ripple-wavelength at their beginning, which becomes:
\begin{equation} \label{lks}
\hat \lambda \sim 2 \pi \sqrt{\frac{2 {\mathcal K }}{\nu + s }} \;\;\;\;.
\end{equation}

\section{Exponential/power-law growth and critical roughness}

In metals, when the Erlich-Schwoebel barrier is active, there is an important non-linear contribution in the current of mobile atoms which becomes sizable when the roughness becomes sufficiently large and therefore $\left< (\alpha_d {\mathbf \nabla} h)^2 \right> \sim 1$ (see Eq.\ref{J1}) where the average is over all the surface positions.
We observe numerically that this changes the law of growth of ripples: from an exponential to a power-law behavior.
This effect is shown in Fig.\ref{f.3}.
Numerically, all the computed exponents follow in the range between $0.65$ and   $0.85$.
The theoretical evaluation of this exponent is under current investigation.
In this regime the ripple-wavelength tends also to grow with time.
Experimentally, power law growth of the roughness and growth of characteristic wavelengths were observed in erosive sputtering on Ag(001) \cite{Valbusa02}. 

In semiconductors or glasses, when no Erlich-Schwoebel barrier is present, it is physically intuitive that the exponential growth of the surface roughness (which is a characteristic of the beginning of the surface instability) cannot continue indefinitely. 
Indeed, from the expression $ R({\mathbf r },t) = R_0 + \hat R_1 \exp(i \omega t + i {\mathbf k } {\mathbf r })$, which we used to derive Eq.\ref{wave}, we can immediately observe that when $\hat R_1 > R_0 = (1-\phi) \eta/ \gamma $, the amount of mobile atoms might become negative.
Since a negative amount of atoms is physically impossible, the process of exponential roughness-growth described above must necessarily finish around a critical roughness given by:
\begin{equation}\label{Wc}
W_c \sim (1-\phi) \frac{\eta}{\gamma} \;\;\;.
\end{equation} 
This behavior is confirmed by numerical solutions of Eq.\ref{E1} and it is expected to be observable in semiconductors and glasses after sufficiently long times.

\section{Conclusions}

We have shown that the same theoretical approach introduced to describe the formation of aeolian sand ripples can be conveniently applied to the study of the formation of periodic structures on surfaces under ion sputterning.
Although the two phenomena are rather different, they can be described within the same conceptual framework by using rather general ideas that relate mobility, excavation and adsorption rates with the surface shape and orientation.

We have obtained general expressions for the ripples wavelength in term of the system parameters.
It has been shown that in some particular cases such a solution coincide with the ones already known in the literature for sand dunes and surface instability \cite{Bra88,KPZ,Par99,Terz98,Val99}.
We have discussed the effect of the Erlich-Schwoebel barriers and compared the result with numerical solutions.
We pointed out the Erlich-Schwoebel barrier can be responsible for a dramatic change in the system dynamics: from the exponential growth to a power law.
Finally, we have demonstrated that the occurrence of a critical roughness is predictable within the present theoretical framework.

It should be noted that the main purpose of this paper is to point out a relevant example of universality: two processes which have completely different scales present a dynamical evolution which obeys to the same geometrical constraints and thus can be described by using the same phenomenological model.
On the other hand, we must observe that the class of solutions of Eq.\ref{E1} is rich and complex - even in the linear approximation.
Exhaustive, systematic studies of the classes of solutions of this equation and their dependence on the set of parameters will be the subject of future studies and publications.


\appendix

\section{Fourier transform of the linearized equation} \label{A1}

By substituting Eqs.\ref{R1h1}, \ref{J1}, \ref{Gad} and \ref{Gex} into Eq.\ref{E1} and by neglecting the second order terms (in $R_1$ and $h_1$), we obtain the following linearized equation:
\begin{eqnarray} \label{EqStab}
\frac{\partial h_1}{\partial t} 
&=&  
   \gamma R_1 
- \left[ {\mathbf v}_1 - (1-\phi){\mathbf v}_2 \right] {\mathbf \nabla} h_1 
- 
\nonumber \\
& &
\left[ {\mathcal D }_1 - (1-\phi){\mathcal D }_2 \right] {\mathbf \nabla}^2 h_1 
\nonumber \\
\frac{\partial R_1}{\partial t} 
&=& 
- \gamma R_1
- {\mathbf v }  {\mathbf \nabla } R_1 
+ {\mathcal D } {\mathbf \nabla }^2 R_1 +
\nonumber \\
&& (1-\phi)\Big[ ({\mathbf v}_1-{\mathbf v}_2)   {\mathbf \nabla}  h_1 +   
\nonumber \\
&& ({\mathcal D }_1 -{\mathcal D }_2 - s_1    ) {\mathbf \nabla}^2 h_1
- {\mathcal K }_1 {\mathbf \nabla}^4 h_1 \Big]
\;\;\;\;.
\end{eqnarray}

A Fourier analysis of Eq.\ref{EqStab} leads to
\begin{eqnarray} \label{EqFou}
& &
\gamma  \hat R_1 
-\Big\{
i \omega 
+  
i {\mathbf k } \left[ {\mathbf v }_1 - (1-\phi) {\mathbf v }_2 \right] -
\nonumber \\
& & 
k^2 \left[ {\mathcal D }_1 -(1-\phi) {\mathcal D }_2 \right]
\Big\} \hat h_1 = 0
\nonumber \\
& &
\left[ 
i \omega + 
\gamma 
+ i {\mathbf k } {\mathbf v } 
+  k^2 {\mathcal D } 
\right]  \hat R_1 
-
(1-\phi)
\Big[
  i {\mathbf k } ({\mathbf v }_1 -{\mathbf v }_2) 
-
\nonumber \\
& &
k^2  \left({\mathcal D }_1 - {\mathcal D }_2 - s_1 \right)
- k^4 {\mathcal K }_1
\Big] \hat h_1 = 0
\;\;\;,
\end{eqnarray}
with $\hat R_1$ and $\hat h_1  $ the Fourier components of $R_1$ and $h_1$ respectively.
This equation is a simple linear equation in two variables. 
It admits a non-trivial solution when the determinant of the coefficients is equal to zero.
This leads to Eq.\ref{Deter}.

\section{Exact solutions} \label{A2}

Analytical expressions for the value of $\mathbf k$ at which $Im(\omega)=0$ (${\mathbf k}^*$) can be calculated from Eq.\ref{solut} in some special cases. 

In particular, when $\phi=0$, $s_1=0$, ${\mathcal K }_1=0$ and ${\mathcal D }=0$, we obtain
\begin{equation}\label{a0}
k^*
=
\sqrt{
\frac{
 \gamma ({v }_1 -{ v }_2)   }{ 
(v - { v }_1 + { v }_2)
 ({\mathcal D }_2 - {\mathcal D }_1) 
} 
}  
\;\;\;,
\end{equation}
where $v$, $v_1$ and $v_2$ are the components of ${\mathbf v}$, ${\mathbf v}_1$ and ${\mathbf v}_2$ in the direction of ${\mathbf k }^*$.

On the other hand when, ${\mathcal K }_1$, $s_1$, ${\mathcal D }_1$ and ${\mathcal D }_2=0$, we find 
\begin{equation}\label{a3}
k^*
=
\sqrt{
\frac{
\gamma ({v } - \phi { v }_1)
}
{
\mathcal D ({v }_1 - { v } -(1-\phi) { v }_2) 
}} \;\;\;.
\end{equation}

The effect of the deterministic diffusion induced by the chemical potential can be studied from the solution  \begin{equation}\label{a1}
k^*
=
\sqrt{
\frac{\phi \gamma {\mathcal D }_1}{ 
(1-\phi) \gamma {\mathcal K }_1  - {\mathcal D }({\mathcal D }_1 - (1-\phi) {\mathcal D }_2) } }  \;\;\;,
\end{equation}
which holds when ${\mathbf v }=0$, ${\mathbf v }_1=0$, ${\mathbf v }_2=0$, $s_1=0$ and ${\mathcal D }-{\mathcal D }_1 + (1-\phi) {\mathcal D }_2 >0$.

Whereas, when ${\mathbf v }_1$, ${\mathbf v }_2$, ${\mathcal D }_1$ and ${\mathcal D }_2=0$, we find 
\begin{equation}\label{a2}
k^*
=
\sqrt{
\frac{ s_1 }{ 
{\mathcal K }_1  } }  \;\;\;,
\end{equation}
\noindent
which implies that the uphill current due to the Erlich-Schwoebel barrier can generate instability even when the shape-dependent erosion and recombination terms are inactive.

\section{Numerical Solutions} \label{A.sim}

The numerical solutions of Eq.\ref{E1} presented in this paper and in particular the ones shown in Figs.\ref{f.1} and \ref{f.3} have been performed as follows.
We considered a one-dimensional flat substrate ($h(x,0)=h0$) of length $L$, with periodic boundary conditions. 
An infinitesimal quantity of mobile atoms were added randomly to the substrate (with $0 < r(x,0) < L 10^{-10}$).
We then computed the profile-evolution using Eq.\ref{E1} with the derivative substituted with finite differences. 
To this purpose, substrate has been divided into $N$ discrete points.
The -adimensional- time indicated in Fig.\ref{f.3} is the number of numerical steps.
The height is in unit of $L/N$ and the roughness is defined as $w(t,L)=\left< [h(x',t)- \left< h(x,t) \right>_x]^2 \right>^{1/2}_{x'}$ (see, for instance, \cite{Csa92}).

Several computations with a number of points equal to $N=100$, 200 and 300 (the one published have $N=300$) have been performed to verify the effect of boundary and discretization.
Moreover, simulations with no periodic boundary conditions and with the sputtering term (Eq.\ref{Gex}) applied only to a central mask, have also been performed obtaining very similar results.
The robustness of the present approach has been verified varying the parameters, the time steps, the initial roughness of the substrate, etc.
Comparable results have been always found but, we must stress that, under some conditions, numerical instabilities (in particular small surface-deformations with $\lambda \sim L/N$) can be trigged on depending on the protocol utilized.

%

\begin{thebibliography}{99}
\bibitem{typical}
Although this is the typical behavior for $Im(\omega({\mathbf k}))$, some other more complex behaviors can be observed for special ranges of the parametes.
\end{thebibliography}
%



\end{document}